\documentclass[useAMS,usenatbib]{mn2e}

\usepackage{caption}
\usepackage{natbib}
\usepackage{graphicx}
\usepackage{textcomp}
\usepackage{latexsym}
\usepackage{varioref}
\usepackage{xspace}
\usepackage{makeidx}
\usepackage{verbatim}
\usepackage{tabularx}
\usepackage{epstopdf}
\usepackage{amsmath}
\usepackage{array}
\usepackage{rotating}
\usepackage{hyperref}
\usepackage{float}

\def\Kepler{\textit{Kepler}}

\title[Physical model for flickering in CVs]
{A physical model for the flickering variability in cataclysmic variables}
\author[S. Scaringi]
{Simone Scaringi$^{1}$\thanks{E-mail: simone.scaringi@ster.kuleuven.be}\\ 
$^{1}$Instituut voor Sterrenkunde, K.U. Leuven, Celestijnenlaan 200D, B-3001 Leuven, Belgium \\
}
\begin{document} 

\date{}

\pagerange{\pageref{firstpage}--\pageref{lastpage}} \pubyear{2013}

\maketitle

\label{firstpage}

\begin{abstract}
Aperiodic broad-band variability (also known as flickering) is observed throughout all types of accreting compact objects. Many statistical properties of this variability can be naturally explained with the fluctuating accretion disk model, where variations in the mass-transfer rate through the disk are modulated on the local viscous timescale and propagate towards the central compact object. Here, a recently developed implementation of the model is applied for the first time to the time-averaged, high-frequency variability of a cataclysmic variable star (MV Lyrae) observed with the \Kepler\ satellite. A qualitatively good fit to the data is achieved, suggesting the presence of geometrically thick inner flow with large viscosity parameter, extending from $\sim0.12R_{\odot}$ all the way to the white dwarf surface. A simple spectral model of the system suggests that the geometrically thick component would not contribute much to the observed optical flux originating from the geometrically thin outer disk. Instead, X-ray reprocessing from the geometrically thick disk onto the thin disk is proposed as a mechanism to explain the observed variability. Similar flows are also deduced in accreting neutron stars/black holes (X-ray binaries) and Active Galactic Nuclei. Additionally, eclipse mapping studies of cataclysmic variables also seem to suggest the presence of a geometrically extended flow towards the inner-edges of the accretion disk. The fluctuating accretion disk model applied here is encouraging in understanding the origin of flickering in cataclysmic variables, as well as in X-ray binaries and Active Galactic Nuclei, by providing a unifying scheme by which to explain the observed broad-band variability features observed throughout all compact accreting systems.

\end{abstract}


\section{Introduction}
Aperiodic broad-band variability (also referred to as flickering) is observed throughout all classes of compact interacting binary systems consisting of a late-type star transferring material onto a compact object such as a white dwarf (WD), neutron star (NS) or black hole (BH). For low-mass systems, mass-loss from the secondary star occurs through the L1 point, after which it forms an accretion disk surrounding the compact object. As angular momentum is transported outwards, material will approach the inner-most regions of the disk, and in the absence of strong magnetic fields, eventually accrete onto the compact object. Most of the radiation produced by the accretion process is emitted through the release of gravitational potential energy. In the case of accreting BHs and NSs the bulk emission takes the form of X-ray radiation (and are thus called X-ray binaries or XRBs), whilst for accreting WDs (also referred to cataclysmic variables or CVs) most of the emission takes the form of optical/ultraviolet 
radiation. This difference can be attributed to the mass and size of the accreting compact objects (or event horizon for BHs), which in the case of accreting BHs/NS allows material to fall much deeper within the gravitational potential well as opposed to accreting WDs (which in contrast have a radius of $\approx3$ orders of magnitude larger than NSs or BH event horizons for a similar mass).

Theoretically and observationally, XRBs and CVs share many similarities between them. Both are observed to display large-amplitude outbursts (\citealt{wood11,cannizzo12,odon}) which are theoretically explained by the thermal-viscous instability (\citealt{ss_73,lasota01}). Additionally, XRBs have been observed to undergo so-called state transitions, where systems are observed to display an evolution in their spectral state in conjunction with their luminosities. At their lowest X-ray luminosities, XRBs display a hard X-ray spectrum (photon index $\Gamma<2$) with a weak accretion disk component. This transitions to a soft, disk-dominated, spectrum during the X-ray outburst (\citealt{rem_mc}). The transition also occurs in conjunction with radio emission associated to the launch of relativistic jets (\citealt{fender}). Similarly, also CVs are though to display a similar state transition in conjunction with jet launching during outbursts, which has been observed in the prototypical CV system SS Cyg (\citealt{
koerding}). 

In addition to the above similarities, XRBs and CVs possess similar intrinsic aperiodic broad-band noise variability (flickering) properties. Both types of systems display a linear relation between the absolute rms (root-mean-square) variability integrated between any two timescales and the flux average (the so-called rms-flux relation; \citealt[][Scaringi et al. 2012a]{uttley1,uttley2}). This observation seems to hold over all timescales, and rules out additive shot noise models (\citealt{terrell,weiss}) in favour of models where variability on different timescales is coupled together multiplicativley. Additionally, the broad-band variability of both CVs and XRBs display single or multiple quasi-periodic oscillations (QPOs, \citealt{mauche,warner_qpo,pretorius_qpo,vanderklis_aper,vanderklis96,belloni1,belloni2}) in conjunction with a high frequency break. These features can be phonologically modelled by a combination of Lorentzian-shaped functions. The main difference between the QPOs and high frequency 
breaks observed is that the characteristic frequencies\footnote{The characteristic frequency is defined as $\Delta\nu^{2}_
{peak}=\Delta\nu^2 + \Delta\nu^2_0$, where $\Delta\nu^2$ and $\Delta\nu_0$ are the half width at half maximum and centroid respectively.} are $\approx3$ orders of magnitude lower in temporal frequency for CVs when compared to XRBs (\citealt{belloni2,revn1,revn2}). The variability of both XRBs and CVs is also highly coherent across different energy bands/wavelength ranges, and Fourier dependent time-lags are observed in both types of systems (\citealt[][Cassatella et al. 2012a,b]{vaughan,nowak,fabian09,uttley11,scaringi_lags,demarco}). 

The physical process governing the observed timing properties in XRBs (and possibly CVs) is often attributed to propagating fluctuations in mass accretion rate within the accretion disk (\citealt{lyub,kotov,AU06}). In this model, the observed variability is associated with modulations in the effective viscosity of the accretion flow at different radii. More specifically, the model assumes that, at each radius, the viscosity -- and hence accretion rate -- fluctuates on a local viscous time scale around the mean accretion rate, whose value is set by what is passed inwards (again on the viscous time scale) from larger radii. The overall variability of the accretion rate in the innermost disk regions (which is then observed through X-ray radiation in XRBs and optical/UV radiation in CVs) is therefore effectively the product of all the fluctuations produced at all radii. In this respect, \cite{ID11,ID12} (hereafter ID11 and ID12 respectively) have developed a numerical model which allows to fit the PSD properties 
observed in XRBs by implementing the fluctuating accretion disk model in the context of a hot inner flow. The model allows, under certain assumptions, to fit the PSD and infer disk parameters such as the inner/outer disk radius, mass of the compact object, and the strength of the effective viscosity that drives the mass transfer through the disk. 

Recently \cite{IK13} (hereafter IK13) have derived exact and analytical expressions for the fluctuating accretion disk model, which allows to produce PSDs for any combination of model parameters without resorting to Monte Carlo simulations, thus drastically reducing the computation time required to compute such models. Driven by the growing similarities between the broad-band variability properties between CVs and XRBs, and the recent analytical treatment of the fluctuating accretion disk model of IK13, this Paper is concerned with the application of such model (developed for XRBs) to the broad-band PSD of the accreting WD MV Lyrae. MV Lyrae has been the subject of many recent studies thanks to the short cadence data obtained by the \Kepler\ satellite. Specifically, Scaringi et al. 2012a \nocite{scaringi1} (hereafter Paper~I) have shown how the high-frequency flicker noise of this system displays the linear-rms-flux relation as observed in XRBs and AGN, whilst Scaringi et al. 2012b \nocite{scaringi2} (
hereafter Paper~II) studied the broad-band variability behaviour of this system, noting the many phenomenological similarities between the PSD properties of this system to XRBs and AGN. This work is concerned with applying the physical model of IK13 to the updated dataset (including the most recent \Kepler\ observations) in order to gain physical insight into the driving variability mechanism at high-frequencies of MV Lyrae in particular, and CVs in general. 

\begin{figure*}
\includegraphics[width=1\textwidth]{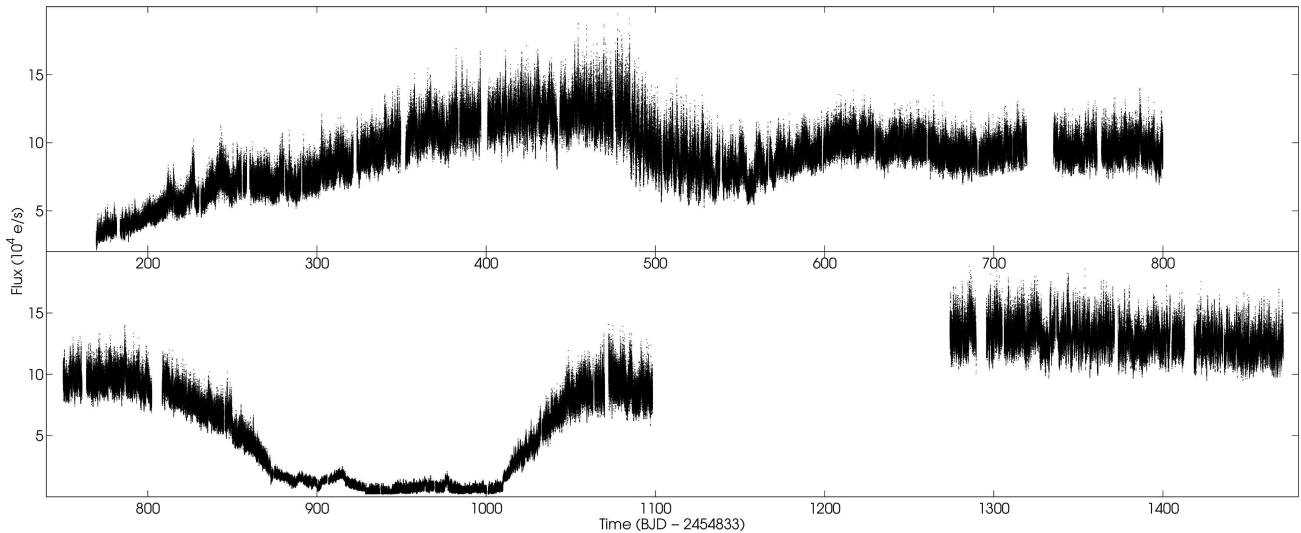}
\caption{MV~Lyrae light curve obtained by the \Kepler\ satellite in short cadence mode (58.8 s).}
\label{fig:1}
\end{figure*}

\section{Observations}\label{sec:obs}

The MV Lyrae lightcurve used in this work has been obtained with the \Kepler\ satellite during 15 quarters of \Kepler\ operations. The first 633 days of data are the same as that used in Paper~I and Paper~II. The lightcurve has been obtained from the Mikulski Archive for Space Telescopes (MAST) in reduced and calibrated form after being run through the standard data reduction pipeline (\citealt{jenkins}), and spans an interval of 1301.6 days. Only the Single Aperture Photometry (SAP) lightcurve is considered. Data gaps occasionally occur due to \Kepler\ entering anomalous safe modes (see Kepler Data Release Notes 20; \citealt{DRN20}). During such events no data are recorded and for a few days following these events the data are always correlated due to the spacecraft not being in thermal equilibrium. Here, no attempt is made to correct these artifacts, but are simply removed them from the light curve. Fig.\ref{fig:1} shows the short cadence (58.8 seconds; \citealt{gilliand}), barycentre corrected, lightcurve 
for MV Lyrae used in this work.

A \textit{phenomenological} analysis of the broad-band PSD of MV Lyrae has already been presented in Paper~II. In that work, the authors fitted the PSD with a combination of 4 Lorentzian-shaped functions plus a power law, in a similar way to what is fitted to the PSDs of XRBs. The authors showed how the PSD of MV Lyrae is in many ways similar to those observed in XRBs, although for MV Lyrae the observed characteristic frequencies seem to be scaled down $\approx$ 3 orders of magnitude in temporal frequency. Additionally, the authors have demonstrated how the fitted characteristic frequencies vary during the observation. This work is concerned with fitting the \textit{physical} model of IK13, developed for X-ray binaries, to the PSD of MV Lyrae. Because of the time-varying complexity of the broad-band PSD shape of MV Lyrae, only the time-averaged, high-frequency break/Lorentzian ($L_1$ in Paper~II) will be considered. This Lorentzian is the least frequency-varying Lorentzian observed in MV Lyrae. In a later 
work, a time-varying fit to the whole PSD will be attempted.

\begin{figure}
\includegraphics[width=0.5\textwidth]{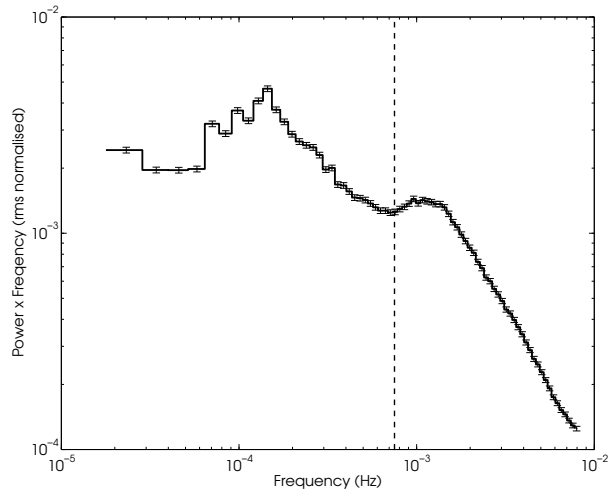}
\caption{Time-averaged PSD of MV~Lyrae. The PSD has been obtained by averaging 223 5-day PSD segments after subtracting for white noise.}
\label{fig:2}
\end{figure}

In order to obtain the time-averaged PSD of MV Lyrae, the lightcurve in Fig. \ref{fig:1} has been split into 260 non-overlapping segments, each covering 5 days (similarly to Paper~II). Fast-Fourier transforms (FFT) of 223 segments are then computed, ignoring 37 segment due to data gaps. The rms normalisation of \cite{miyamoto} and \cite{belloni2} is then applied to each FFT, so that the square root of the integrated PSD power over a specific frequency range yields the rms variability. All segments are then averaged together and re-binned geometrically, using a re-binning constant of 1.05 (\citealt{klis89,IK13}), yielding over $263$ measurements per frequency bin. The high number of measurements ensures that the statistical errors on the PSD have converged to the Gaussian limit for each frequency bin, and thus the use of $\chi^2$ as a fit statistic is appropriate. The resulting time-averaged PSD is shown in Fig. \ref{fig:2}. Given that this work is only concerned with fitting the high frequency Lorentzian/
break with the physical model of IK13, only frequencies above $7.5\times10^{-4}$ Hz are considered (marked with the vertical dashed line in Fig. \ref{fig:2}). Below this frequency the variability power in MV Lyrae begins to rise again, and a more detailed physical model is required in order to take this into account. In a later work, an attempt to fit the whole available frequency range (taking into account the 4 Lorentzians discussed in Paper~II) in a time-dependant fashion will be assessed. 

\section{The Model}\label{sec:model}

The model used to fit the data in this work has been described in detail in ID11, ID12 and IK13.  A brief description of the model is also given here together with the relevant modifications. The model consists of fluctuations in the mass accretion rate which propagate towards the central compact object (\citealt{lyub,kotov,AU06}). The accretion flow extends between an outer and inner radius, $r_o$ and $r_i$ respectively. The flow is split into $N$ rings, each centred at $r_n$, with equal logarithmic spacing such that $dr_n/r_n=$ constant. Fluctuations in the mass-transfer rate at each ring are stochastic, but have a well defined power spectrum producing the maximum variability power at the local viscous frequency, and can be expressed as
\begin{equation}
|A_n(\nu)|^2 =
\frac{\sigma^2}{T\pi}\frac{\Delta\nu_n}{(\Delta\nu_n)^2+\nu^2},
\label{eqn:1}
\end{equation}
where $\Delta\nu_n=f_{visc}=1/t_{visc}$, and $\sigma^2$ and $T$ are the variance and duration of the time series respectively (using the same notation as IK13). The model assumes that $\sigma$ is constant for all rings, such that $\sigma=F_{var}/N_{dec}$, where $N_{dec}$ is the number of rings per radial decade and $F_{var}$ is the fractional variability. This prescription provides the physical assumption that each ring in the flow generates the same amount of variability power (\citealt{beckwith}). 
 
The \cite{ss_73} accretion disk model provides the viscosity prescription 
\begin{equation}
f_{visc}(r_n) = 
\alpha(h/r)^2f_{dyn}(r_n),
\label{equ:2}
\end{equation}
where $f_{dyn}(r_n)$ is the dynamical (Keplerian) frequency at $r_n$, and $\alpha$ and $(h/r)$ are the viscosity parameter and flow semi-thickness respectively. In the model prescription of ID11, $\alpha(h/r)^2$ takes the form of a power law function of radius, which has been found from fitting the QPO-break relation of accreting black holes. On the other hand, ID12 and IK13 relate the viscous frequency to the disk surface density (guided from the GRMHD simulations of \citealt{fragile07,fragile09_1,fragile09_2}). For CVs it is also possible that the viscous frequency is related to the disk surface density, which would provide a power law radius dependence for $\alpha(h/r)^2$. However, a functional form of this power law dependence has not been established for CVs, and $\alpha(h/r)^2$ is kept as a constant in this work for simplicity. With this prescription, constant $dr_n/r_n$ also implies $df_n/f_n$, which \cite{AU06} show is a feature required to produce the linear rms-flux relation.

The transformation from mass accretion rate fluctuations $\dot{m}(r_n)$ to an actual lightcurve requires an emissivity $\epsilon(r_n)$, such that the luminosity from each ring is $\propto \dot{m}(r_n)\epsilon(r_n)r_ndr_n$. With this prescription, $\epsilon(r_n)\propto r_n^{-\gamma} b(r_n)$, where $\gamma$ is the emissivity index and $b(r_n)$ is a boundary condition. ID11 discuss two boundary conditions: a stress-free $b(r_n)=3(1-\sqrt{r_n/r_i})$ and a stressed $b(r_n)=1$ condition. For a Newtonian thin disk, the bolometric emission will have $\gamma=3$. This work employs a stressed boundary condition, since this seems more realistic for accreting WDs where the inner disk can possess torques due to the possible WD rotation and/or magnetic fields, however a stress-free boundary condition is also addressed in Sec. \ref{sec:results}. Furthermore, because the observed data is monochromatic, $\gamma$ is set as a free parameter. 

The inner regions of accretion disks surrounding compact objects are expected to produce a harder spectrum when compared to the cooler outer regions. This in turn implies a steeper emissivity index $\gamma$ for shorter wavelengths. It is this feature that allows the fluctuating accretion disk model to also produce hard Fourier dependent time-lags (\citealt{AU06,kotov}). This work however only considers the single band PSD in order to qualitatively fit the MV Lyrae \Kepler\ lightcurve. Future work will also consider fitting the Fourier-dependent time-lags observed for this system (as well as LU Cam, see \citealt{scaringi_lags}). However, modifications to the model will be required as the Fourier lags observed in MV Lyrae are soft (red photons lagging blue ones) rather than hard. Similar soft lags have also been observed in XRBs and AGN (\citealt[][Cassatella et al. 2012a,b]{demarco,fabian09}), and are explained as reprocessing of X-ray photons produced close to the compact object (by the Comptonised component 
observed as a power law in hard X-ray spectra) on to the accretion disc, either as a thermal blackbody (e.g. X-ray heating from the disc) or, in the case of AGN, as an additional soft photoionized reflection component. In CVs it is also possible that a similar mechanism is at play, but the reprocessing must then occur on a different timescale (possibly the thermal timescale; \citealt{scaringi_lags}).

For accreting black holes the inner disk radius is set to be larger than the last stable circular orbit, which ID12 and IK13 set to a fixed value of $3.3R_g$ (where $R_g=GM/c^2$) when fitting the PSD of XRBs. Here also the inner hot flow radius is assumed to be fixed, and the intuitive assumption that it reaches the WD surface is employed. Given the fitted mass of $0.73M_\odot$ for the WD in MV Lyrae (\citealt{hoard,linnell}), it is possible to use the mass-radius relation for WD (\citealt{HS61,nauenberg}), and set $r_{in}=r_{WD}=0.0105R_\odot$. 

\begin{figure}
\includegraphics[width=0.45\textwidth]{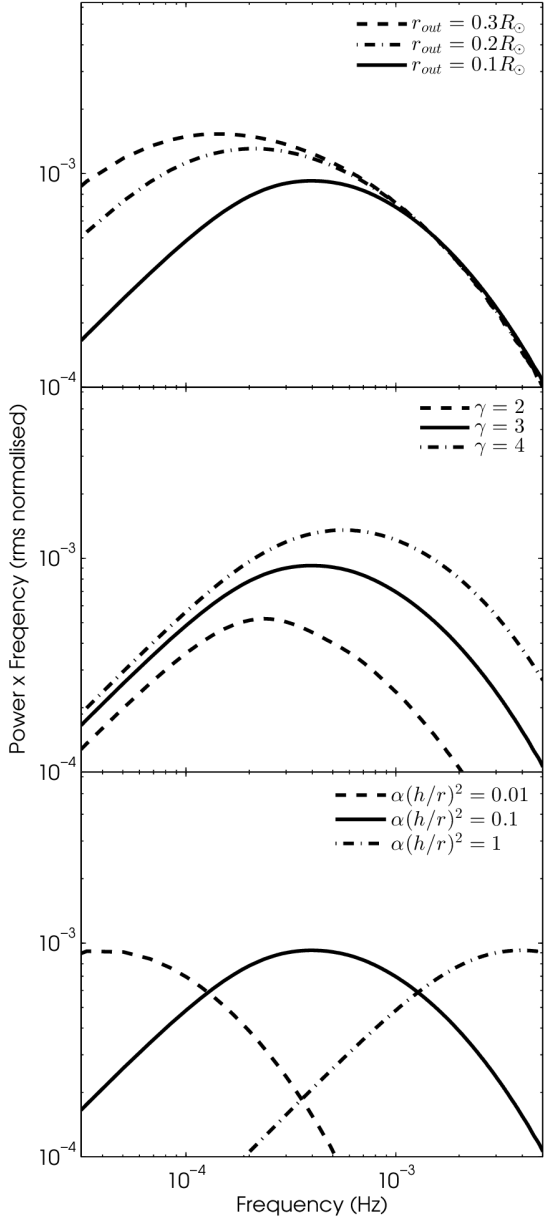}
\caption{Power spectra generated with the model described in Section \ref{sec:model}. The solid line in all three panels is the same, with $R_{out}=0.1R_{\odot}$, $\gamma=3$, $\alpha(h/r)^2=0.1$ and $F_{var}=0.1$. the top panel shows the effect of varying the outer disk radius $r_{out}$. The middle panel shows the effect of varying the emissivity index $\gamma$, whilst the bottom panel shows the effect of varying the $\alpha(h/r)^2=0.1$ parameter.}
\label{fig:3}
\end{figure}

Given the above model description, the analytical method of IK13 is employed to produce the PSDs for CVs. This method allows to produce PSD models for a given set of parameters by convolving the variability power generated at each disk ring in the Fourier domain rather than having to simulate time series. In the prescription used here there are four free parameters, namely $r_{out}$, $\gamma$, $F_{var}$, and $\alpha(h/r)^2$. Fig.\ref{fig:3} shows the effect on the obtained PSD shapes by varying these parameters. The top panel of Fig. \ref{fig:3} keeps all parameters fixed while varying the outer disk radius, which has the effect of producing more variability power at low frequencies for disks extending to larger radii. The middle panel varies the emissivity index $\gamma$ whilst keeping all parameters fixed, and has the effect pushing the high frequency break to higher frequencies for larger values of $\gamma$. The bottom panel varies $\alpha(h/r)^2$, and this has the effect of shifting the whole PSD to 
higher frequencies for larger values of $\alpha(h/r)^2$. All PSD models in Fig. \ref{fig:3} have $F_{var}=0.1$, and the the thick line in all panels is the same with $r_{out}=0.1R_{\odot}$, $\gamma=3$ and $\alpha(h/r)^2=0.1$.

\section{Results}\label{sec:results}
\begin{figure}
\includegraphics[width=0.5\textwidth]{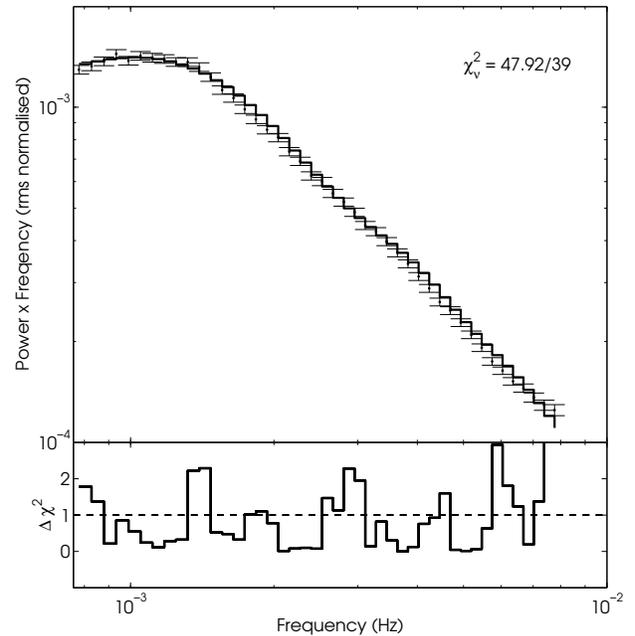}
\caption{White noise subtracted PSD of MV Lyrae with the best model fit using the parameters shown in Table \ref{tab:1} and the resulting $\Delta\chi^2$ from the fit.}
\label{fig:4}
\end{figure}

In order to obtain a best-fit to the time-averaged high-frequency data of Fig. \ref{fig:2} using the model described in Section \ref{sec:model}, $\chi^2$ is used as an objective function to minimise. Given the $\chi^2$ landscape might have local minima as well as a global one, a procedure by which the objective function is minimised from different starting points (implemented in MATLAB as Multistart, \citealt{multistart}) is employed. All initial starting points will converge to some local minima, and the best fit of all these is taken as the global one. In order to ensure the parameter space is fully covered, 5000 initial starting points are used, and models with $N=30$ number of rings are computed. Fig. \ref{fig:4} shows the best-fit result plotted in frequency $\times$ power, with the contributions to $\chi^2$ plotted underneath. The best fit parameters, shown in Table \ref{tab:1}, yield a minimum $\chi^2_{\nu}=47.92/39=1.23$, and suggest the presence of an inner disk extending from $r_{out}=0.117$ all 
the way to the WD surface with a relatively large $\alpha(h/r)^2$. A similar fitting procedure has been employed using a stress-free boundary condition (see Sec. \ref{sec:model}). The minimum $\chi^2_{\nu}=76.39/39=1.96$ is however much larger than the one found for a stressed boundary condition.

\begin{figure*}
\includegraphics[width=1\textwidth]{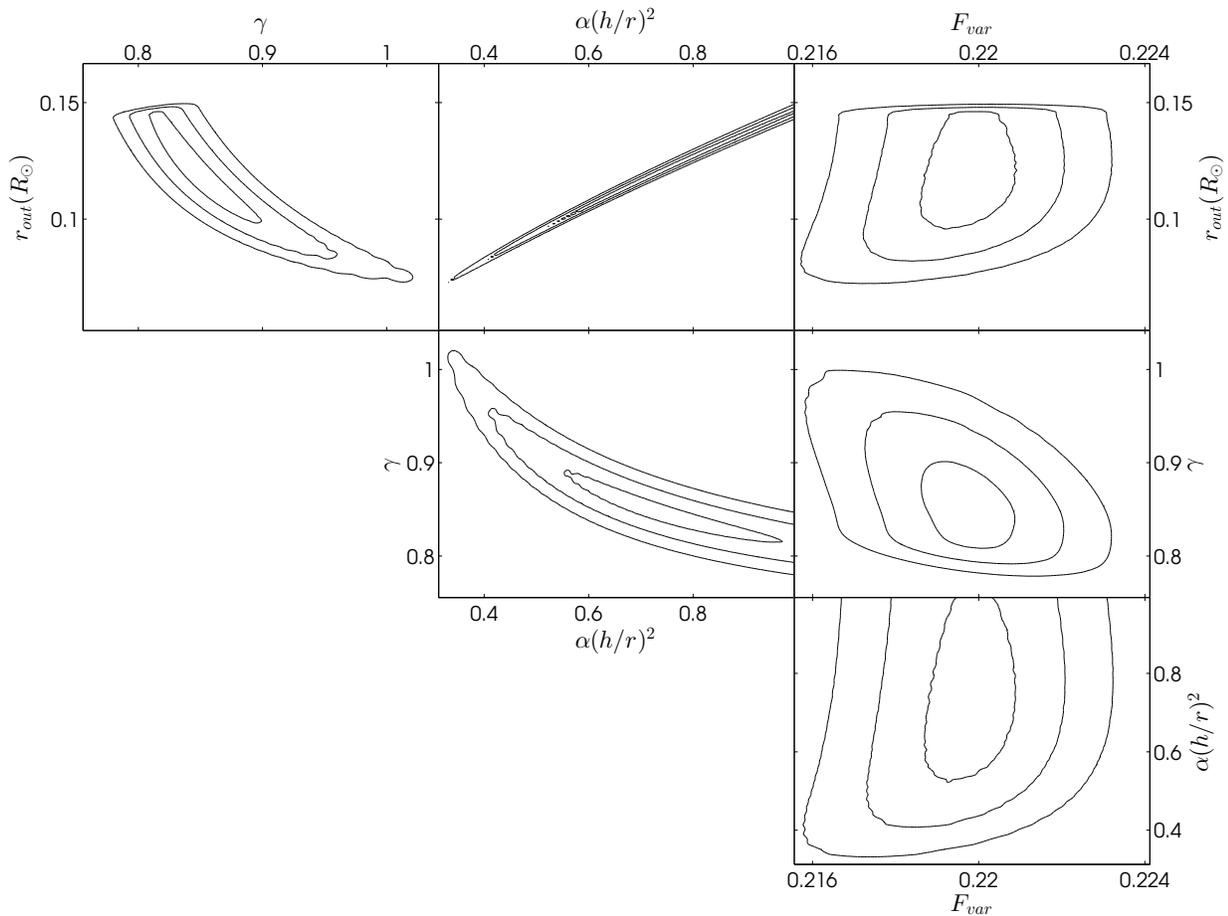}
\caption{Confidence contour regions corresponding to $68.3\%$ ($1\sigma$), $95.4\%$ ($2\sigma$) and $99.73\%$ ($3\sigma$) for all four free parameters shown in Table \ref{tab:1}. Each panel plots confidence contours on 2 parameters allowing the other 2 to be free and take any value. The apparent break at high values of $r_{out}$ in the top-left and top-right panels are the result of the grid search only covering the range $\alpha(h/r)^2\le1$.}
\label{fig:5}
\end{figure*}

In order to obtain confidence intervals on the fitted parameters, a grid search around the best fit has been performed. The $\Delta\chi^2=1$ projections on the four parameter axis then yield the $1\sigma$ confidence intervals (\citealt{NR,lampton}). The $1\sigma$, $2\sigma$ and $3\sigma$ contour intervals are displayed in Fig. \ref{fig:5}. These highlight the correlations between the various fit parameters and in particular show how it can be possible to obtain lower $\alpha(h/r)^2$ by both decreasing $r_{out}$ and increasing the emissivity index $\gamma$. 
 
Assuming the model is correct in explaining the data, the fact that $\chi^2_\nu$ lies above unity can potentially be explained by the fact that the PSD is non-stationary during the 1301.6 days of \Kepler\ observations (see also Paper~II). However, the fairly good fit to the data provides qualitative insight into the physical process generating the high-frequency variability in CVs. In a future work the time dependence of the model parameters will be assessed by analysing the time-dependant PSD of MV Lyrae.

\begin{table}
\renewcommand{\arraystretch}{1.5}
\centering
\begin{tabular}{l l}

 \hline
 \hline
 $M_{WD}(M_{\odot})$  &    $\equiv0.73$       \\ 
 $r_{in}(R_{\odot})$  &    $\equiv0.0105 $      \\   
 $r_{out}(R_{\odot})$   &      $0.117^{+0.029}_{-0.020}$ \\  
 $\gamma$               &   $0.853^{+0.047}_{-0.041}$  \\ 
 $\alpha(h/r)^2$        &   $0.705^{+0.289}_{-0.182}$  \\ 
 $F_{var}$              &     $0.220^{+0.001}_{-0.001}$ \\ 
 \hline
\end{tabular}

\caption{Best-fit parameters with associated ($1\sigma$) errors for the model described in Section \ref{sec:model} to the MV Lyrae PSD shown in Fig. \ref{fig:4}.}
\label{tab:1}
\end{table}

\section{Discussion}\label{sec:discussion}

\begin{figure}
\includegraphics[width=0.5\textwidth]{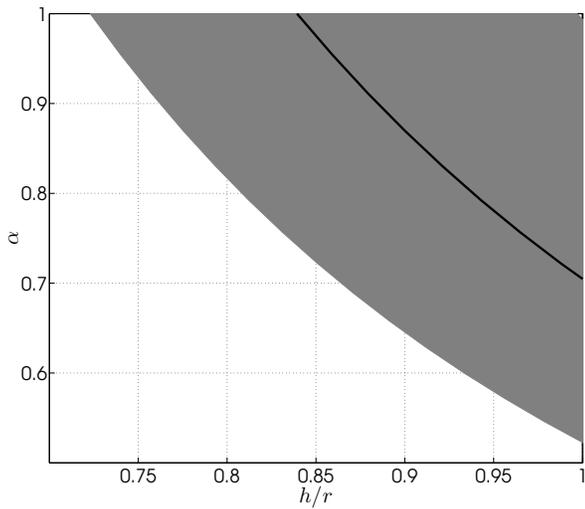}
\caption{Possible $\alpha$ and $h/r$ values for the best-fit parameter in Table \ref{tab:1} (thick black line), together with the $1\sigma$ confidence region (shaded gray region). The obtained suggest the presence of a geometrically thick disk, which in this case would extend from $\sim0.12R_{\odot}$ all the way to the WD surface.}
\label{fig:6}
\end{figure}

The fluctuating accretion disk model developed by ID11, ID12 and IK13 allows to produce PSDs which seek to provide a physical meaning to the flicker noise observed in compact accreting systems. This model has been applied to a CV system for the first time, and a qualitatively good fit has been achieved. The model used here allows to fit for the four parameters $r_{out}$, $\gamma$, $F_{var}$, and $\alpha(h/r)^2$. Of these parameters, $r_{out}$ and $\alpha(h/r)^2$ provide direct insight into the structure of the disk responsible for producing the high-frequency variability observed in CVs. In the MV Lyrae, the obtained values for these parameters seem to suggest that the high-frequency flicker noise is caused by a hot inner flow which extends from $r_{out}\sim0.12R_{\odot}$ all the way to the WD surface. Furthermore, the obtained value for $\alpha(h/r)^2$ seems to be too high to be accommodated by a geometrically thin, optically thick, disk. Fig. \ref{fig:6} shows the possible $\alpha$ and $h/r$ combinations, 
which are in agreement with the results presented in Paper~I and Paper~II. In model prescription used here, $\alpha(h/r)^2$ was set to constant, whilst it would be more physically plausible to allow this parameter to have a radial dependence. Such a modification will be examined in a later work, however it can already be noted that such a modification would still result in the inner disk being relatively thick when fitting the PSD of MV Lyrae. 

The presence of a geometrically thick inner accretion flow in CVs has already been postulated in previous studies using eclipse mapping. For example, \cite{feline05} found that the mid-eclipse colours of the dwarf nova (DN) HT Cas cannot be explained solely by the secondary star or from the outer edges of the accretion disk. Instead, the authors suggest that the mid-eclipse colours originate from a combination of the secondary star with a vertically extended, optically thin, accretion flow. Also using eclipse mapping, \cite{j_wood1} present a map of the brightness temperature distribution of the DN Z Cha. They show how the intensity increases smoothly to a maximum and then flattens towards the inner-disk region. This is in contrast with a steady state, optically thick disk. Here as well the authors suggest the presence of an optically thin component. This result is also further supported by \citealt{j_wood2,groot1,groot2}, which show similar results for V Per, RW Tri and SW Sex. Furthermore, 
\cite{baptista04} demonstrate how the high frequency flicker noise in V2051 Oph is associated to the accretion flow, and discuss that if the flickering is caused by MHD turbulence (\citealt{geertsema}), then the obtained values of both $\alpha$ and $h/r$ would also suggest the presence of an optically thin, geometrically thick disk. For DN in outburst also, \cite{rutten92} find through eclipse mapping that OY Car possesses a bright uneclipsed component, which can possibly be explained by a large scale height disk. Additionally, also studies of the Nova-like (NL) system UU Aqr (which is possibly more similar to MV Lyrae) draw similar conclusions through time-resolved spectroscopy (\citealt{baptista00}). Finally \cite{rutten92_2} and \cite{groot2} have shown how the radial temperature dependence of the accretion disk is flatter than what is expected from a standard geometrically thin disk close to the WD, and suggest an extended boundary layer to explain the results.

Optically thin, geometrically thick, inner-disks are also frequently invoked in order to explain the X-ray spectral-timing properties of XRBs and AGN (\citealt{rem_mc}). The interplay between the cool blackbody-dominated disk component, and the hot Comptonised component is in fact the driving mechanism between state changes in XRBs, and implies the existence of two distinct flows within the accretion disk. The presence of a geometrically thick disk has also been shown to be a requirement in order for the fluctuating accretion disk model to work, and produce the observed rms-flux relations throughout compact accreting systems (\citealt{chur}). The analysis presented here only suggests the presence of a geometrically thick disk at radii smaller than the fitted $r_{out}\sim0.12R_{\odot}$ for MV Lyrae (and possibly other CVs as well). The results do not rule out the presence of a standard geometrically thin, optically thick, disk at larger radii (\citealt{ss_73,lasota01}), or even that a thin disk exists at 
small radii ``sandwiched'' within the thick disk (\citealt{dove}). 

\begin{figure*}
\includegraphics[width=1\textwidth]{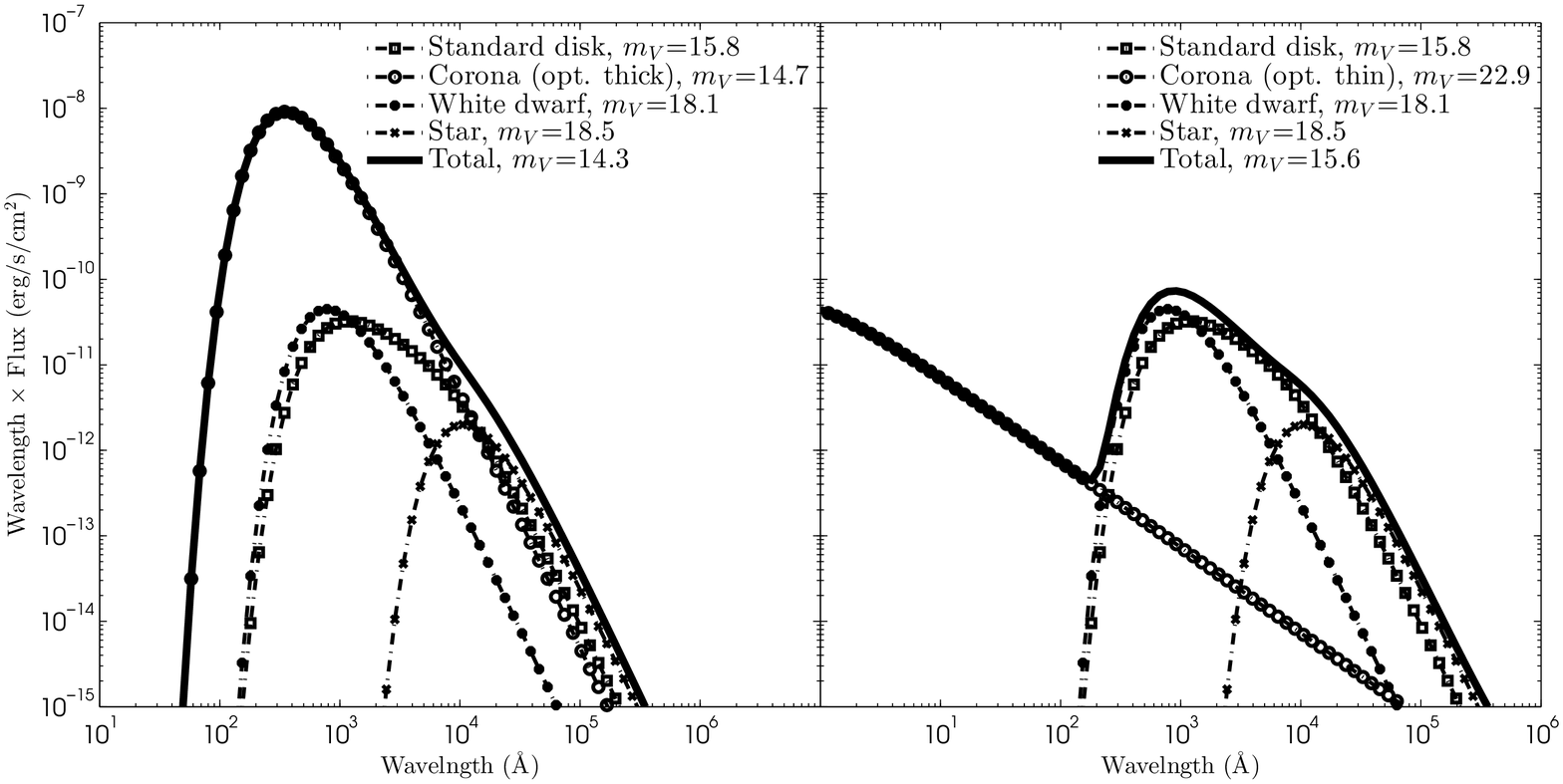}
\caption{Modelled spectral energy distribution and estimated V-band magnitudes for the MV Lyrae system. The accretion disk, secondary star, and WD components are the same in both panels. On the left the geometrically thick disk (corona) is modelled as an optically thick blackbody sphere of radius $0.03R_{\odot}$ with $T=10^{5}$K. On the right the geometrically thick disk is modelled as a Bremsstrahlung emitting plasma with $T=T_{vir}=2.7\times10^8$K and $n_{e}=n_{i}=8\times10^{13}$ cm$^{-3}$. Both models produce similar ROSAT fluxes as those observed for MV Lyrae. However, the bolometric luminosity obtained from the optically thick corona is over 2 orders of magnitudes larger than that inferred from the disk, whilst that obtained from the optically thin corona is comparable at $\approx10^{33}$ erg/s.}
\label{fig:7}
\end{figure*}

At the optical wavelengths observed here most of the emission is though to originate from the outer geometrically thin, optically thick accretion disk. This can be modelled for MV Lyrae by assuming the model parameters reported from ultraviolet spectral fitting of \cite{linnell}, which employ a distance to the system of 505 parsec and a range of mass transfer rates from $1 \times 10^{-10} M_{\odot}$ yr$^{-1}$, during the low luminosity state, to $3 \times 10^{-9} M_{\odot}$ yr$^{-1}$ at high luminosities. The contribution at optical wavelengths from the inferred geometrically thick disk in this work is however not so straight forward to model. The virial temperature around a white dwarf will typically be $\approx10^8$ K, and hence this represents a good initial upper limit of what the temperature of the geometrically thick disk should be. For an undiluted blackbody it can be shown that the optical Johnson V-band magnitude arising from this component would be higher than the high state luminosity of MV Lyrae by $\approx5$ magnitudes. Furthermore, the expected ROSAT X-ray flux would be far higher than the observed flux of MV Lyrae of $\approx4\times 10^{-13}$ erg/s (\citealt{rosat}) by over 9 orders of magnitude. In order to bring the modelled ROSAT flux onto the observed flux, the blackbody temperature would have to be on the order of $10^5$ K. This would then yield an optical magnitude of $m_{V}\approx14.7$, comparable to the observed MV Lyrae intermediate state magnitude. The left panel of Fig.\ref{fig:7} displays the spectral energy distributions obtained from modelling the MV Lyrae system with the model parameters of \cite{linnell} and the inclusion of a $10^{5}$ K blackbody radiating sphere with spherical shape of radius $0.03R_{\odot}$, representing the geometrically thick disk (referred to as Corona). The geometrically thin, optically thick, accretion disk is taken to extend from the L1 point at $\approx0.57R_{\odot}$ all the way to the WD surface, with a mass transfer rate of $3\times10^{-9} M_{\odot}$ yr$^{-1}$. The WD and secondary star are simply modelled as black-bodies with temperatures of $47000$K and $3500$K and radii of $0.01R_{\odot}$ and $0.4R_{\odot}$ respectively (\citealt{linnell}). The inclusion of the $10^5$K, geometrically thick, blackbody component makes the expected ROSAT flux match the observations. However, the integrated bolometric luminosity of this geometrically thick blackbody component would be $\approx10^{35}$ erg/s, over 2 orders of magnitude larger than than inferred from the outer geometrically thin accretion disk.

The modelling of the geometrically thick disk as a blackbody is overly simplistic, and would imply the disk to also be optically thick. A better description would be to assume the disk to be optically thin, in which case Bremsstrahlung emission (rather than blackbody) is more appropriate. The right panel of Fig.\ref{fig:7} displays the same spectral energy distribution obtained from modelling the MV Lyrae system, however replacing the blackbody component representing the corona with a Bremsstrahlung component of the same size. In this case it is possible to obtain the observed ROSAT flux by setting the temperature of the corona to the virial temperature around the WD at $\approx 2.8\times10^8$ K, and setting the electron number density of the plasma equal to that of the ions at $n_{e}\approx8\times10^{13}$ cm$^{-3}$ (similar to electron densities inferred in CVs, eg. \citealt{meyer94}). This scenario is more realistic, and yields a bolometric luminosity of the Bremsstrahlung component comparable to that of the geometrically thin accretion disk at $\approx10^{33}$ erg/s. In this case, the optical emission from this geometrically thick component would not contribute much to the observed optical luminosity. Thus it seems that the only way to account for the large $\alpha(h/r)^2$ inferred in this work is through reprocessing of X-ray radiation onto the geometrically thin, optically thick, disk emitting at optical wavelengths. In this scenario, the geometrically thick disc close to the WD would be a flow all by itself, and it may well be that the inferred $\alpha(h/r)^2$ reflects the state of this geometrically thick disk and not the geometrically thin one.

The possible transition in CVs at $\approx0.12R_{\odot}$ between geometrically thin to geometrically thick has already been suggested. \cite{revn1,revn2} and \cite{balman} have analysed the PSDs of magnetic and non-magnetic CVs from X-ray lightcurves, and inferred a similar disk truncation radius as the one inferred here. However, these authors associate the high-frequency PSD breaks to the dynamical timescale at the inner-most disk edge, and interpret this as the truncation of a geometrically thin disk. This implies a qualitatively different interpretation to the model discussed here. As shown in Fig. \ref{fig:1} and also discussed in ID11 and ID12, the fluctuating accretion disk model produces the high frequency breaks in the PSDs of accreting compact objects through a combination of the inner and outer disk edge viscous timescales and emissivity index $\gamma$, but not necessarily to the dynamical timescale alone. 

MV Lyrae (as well as LU Cam, see \citealt{scaringi_lags}) is known to display Fourier-dependent soft-lags, as well as some XRBs and AGN (\citealt[][Cassatella 2012a,b]{vaughan,nowak,fabian09,uttley11,demarco}). The observed soft-lags can potentially be explained by reprocessing of hard photons produced close to the compact object on to the accretion disc (either via Compton scattering for XRBs, or a slower reprocessing mechanism for CVs). The current model prescription can only produce hard-lags, where blue/hard photons lag the redder/softer ones at the longest observable frequencies as a consequence of fluctuations propagating inwards and thus passing through hotter emitting regions in the disc. In future, it might be possible to incorporate soft-lags into the model prescription and potentially also fit for these in MV Lyrae and other CVs, but this will require a modification to the model to incorporate disk reprocessing to reproduce the observations reported in \cite{scaringi_lags}.

\section{Conclusion}

Aperiodic broad-band variability (flickering) in XRBs and AGN has been modelled in the context of a fluctuating accretion disk. In this model, the observed variability is associated with fluctuations in the mass-transfer rate at different radii, which are governed by the local viscous timescale. As these fluctuations propagate inwards in the disk they couple together, and the observed variability in the inner-most regions of the disk is therefore effectively the product of fluctuations produced at all radii. Here, the recently developed analytical treatment of the model (IK13) has been applied for the first time to a CV, with data obtained with the \Kepler\ satellite. The fitted parameters seem to suggest the presence of a thick inner accretion flow, extending from $\approx0.12R_{\odot}$ all the way to the WD surface, with a relatively high viscosity parameter $\alpha$. Similar flows have already been deduced for XRBs, as well as for CVs using eclipse mapping (\citealt{feline05,j_wood1,j_wood2,baptista04,groot1,groot2}), and the result presented here seems to suggest a similar flow is also present in MV Lyrae.

The qualitatively good fit between the data and the fluctuating accretion disk model is encouraging in understanding the origin of flickering in CVs as well as in XRBs and AGN, and provides a unifying scheme to explain the observed broad-band variability features observed throughout all compact accreting systems with a single model. Future applications of this model will potentially allow to determine whether the seemingly thick inner flow deduced here is also present in other CV systems, and whether the low-frequency variability can be explained by a geometrically thin disk. This would then suggest a transition within the accretion flow similar to what is observed in XRBs. Furthermore, it will also be possible to apply the model in a time-dependent fashion in order to determine changes within the inner thick flow, and whether these changes correlate with other system parameters such as mass transfer rate and/or quiescent/outburst states. 

\section*{Acknowledgements}
This research has made use of NASA's Astrophysics Data System Bibliographic Services. The author acknowledges funding from the FWO Pegasus Marie Curie Fellowship program. Additionally, the author acknowledges the use of the astronomy \& astrophysics package for Matlab (Ofek in prep.). Further acknowledgements go to Christian Knigge, Elmar K{\"o}rding, Thomas Maccarone as well as the anonymous referee for useful and insightful discussions on the topics discussed in this Paper. 

\bibliographystyle{mn2e}
\bibliography{fluc_paper}

\label{lastpage}

\end{document}